\documentclass{aastex631}
\usepackage{soul}

\newcounter{daggerfootnote}
\newcommand*{\daggerfootnote}[1]{%
    \setcounter{daggerfootnote}{\value{footnote}}%
    \renewcommand*{\thefootnote}{\fnsymbol{footnote}}%
    \footnote[2]{#1}%
    \setcounter{footnote}{\value{daggerfootnote}}%
    \renewcommand*{\thefootnote}{\arabic{footnote}}%
    }

\shorttitle{Charge states and abundances in SEPs}
\shortauthors{Lee et al.}

\begin{document}

\def\RSUN{R$_{\sun}$}
\def\kms{$\rm km~s^{-1}$}
\def\cmcube{$\rm cm^{-3$}$}
\def\RSUNs{R$_{\sun}$ }
\def\kmss{$\rm km~s^{-1}$ }
\def\cmcubes{$\rm cm^{-3$}$ }
\def\Qbar{$<\rm Q>$}
\def\Qbars{$<\rm Q>$ }

\title{Solar Energetic Particle Charge States and Abundances with Nonthermal Electrons}

\author{Jin-Yi Lee}
\affiliation{Department of Astronomy and Space Science, Kyung Hee University, Yongin-si, Gyeonggi-do, 17104, Republic of Korea\email{jlee@khu.ac.kr}}

\author{Stephen Kahler$^\dagger$}
\daggerfootnote{Passed away February 5, 2023}
\affiliation{Air Force Research Laboratory, Space Vehicles Directorate, 3550 Aberdeen Ave., Kirtland AFB, NM 87117, USA}

\author{John C. Raymond}
\affiliation{Harvard-Smithsonian Center for Astrophysics, 
Cambridge, MA 02138, USA}

\author{Yuan-Kuen Ko}
\affiliation{Space Science Division, Naval Research Laboratory, Washington, DC, USA}




\begin{abstract}

An important aspect of solar energetic particle (SEP) events is their source populations.
Elemental abundance enhancements of impulsive SEP events, originating in presumed coronal
reconnection episodes, can be fitted to steep power laws of A/Q, 
where A and Q are the atomic mass and ionic charge. 
Since thermal electron energies are enhanced and
nonthermal electron distributions arise in the reconnection process, we might expect that ionic
charge states Q would be increased through ionization interactions with those electron populations
during the acceleration process. The temperature estimated from the SEPs corresponds to the charge state during the acceleration process, while the actual charge state measured in situ may be modified as the SEPs pass through the corona.

We examine whether the temperature estimation from the A/Q would differ with various kappa values 
in a kappa function representing high-energy tail deviating from a Maxwellian velocity distribution. 
We find that the differences in the A/Q between a Maxwellian and an extreme kappa distribution are only about 10-30\%.  We fit power-law enhancement of element abundances as a function of their A/Q with various kappa values. Then, we find that the derived source region temperature is not significantly affected by whether or not the electron velocity distribution deviates from a Maxwellian, i.e., thermal, distribution.  Assuming that electrons are heated in the acceleration region, the agreement of the SEP charge state during acceleration with typical active region temperatures suggests that SEPs are accelerated and leave the acceleration region in a shorter time than the ionization time scale. 

\end{abstract}



\section{Introduction} \label{sec:intro}

Solar energetic (E $\geq$ 3 MeV/nuc) particle (SEP) events are observed in two broad categories,
gradual and impulsive \citep{reames20, reames21}. While the gradual events result from diffusive
shock acceleration (DSA) in coronal and interplanetary shocks usually driven by wide and fast (V$>$1000 km s$^{-1}$) 
coronal mass ejections (CMEs), the impulsive events originate in coronal jets on open
magnetic field lines driven by magnetic reconnection and are characterized by relatively low
intensities and energies (E =$<$20 MeV/nuc).
CME shocks can accelerate particles by the DSA process, and the turbulence and electric fields in the magnetic reconnection regions of solar jets/flares can also produce power-law distributions of energetic particles.  SEP events classified as gradual are attributed to CME shock acceleration, while those classified as impulsive are associated with flare acceleration, $^3$He enhancement and type III radio bursts \citep{reames99}.  Typically, the rise time and duration of gradual events are longer than those of impulsive ones, and protons dominate gradual events while electrons can dominate impulsive ones The source regions of gradual events are widely distributed in solar longitude, while most impulsive ones originate in a narrower range of longitude between 30$^\circ$W and 80$^\circ$W, and gradual SEPs tend to show elemental abundances near normal coronal values, while impulsive SEPs show strong enhancements of Fe and other heavy elements \citep{reames99}. 

To further understand the nature of SEPs, it is important to identify their source regions:  Are they produced in ordinary coronal plasma, in active regions, in coronal plasma enriched in seed particles by earlier events, or in prominence material ejected during CMEs?  One diagnostic for the origin of SEP particles is the elemental composition of the plasma.  A second diagnostic is the charge states of the SEPs.  These are linked, because the acceleration efficiency is expected to depend on the particle's gyroradius, and therefore to scale as $(A/Q)^\alpha$ \citep{caprioli17}, where A is the particle mass, Q is the charge, and $\alpha$ is the power law index, though the enhancement may saturate at large A \citep{hanusch19}.

The earliest measurements of ionization states \citep{luhn1984, luhn1987} have found that gradual events had measured states corresponding to $\sim$~2~MK, 
but impulsive events had $Q_{Fe}\approx 20$ and Ne, Mg, and Si were fully ionized. The high ionization states in impulsive events could be produced in the acceleration region, or they could be produced  by subsequent stripping.  In  the latter case, it was not clear why Ne, Mg, and Si would be enhanced, since they had A/Q=2.  Finally, Q was found to vary with velocity, so the stripping interpretation prevailed \citep[e.g.][]{difabio08}.

Reames and collaborators have explored these diagnostics in a series of papers \citep{reames99b, reames00, reames02, reames14, reames14a, reames16, reames17, reames18a, reames20, reames22a}.  In some events, the average charge state \Qbars of Fe increases with energy, indicating stripping of the Fe ions during acceleration in high density regions \citep{barghouty99, kovaltsov01, difabio08}, but in most events the charge state is independent of energy, at least at moderate energies.  In the events that do not show evidence of stripping during acceleration, it is possible to determine \Qbars  in the source region by fitting $(A/Q)^\alpha$ to the elemental abundances \citep{ellison84, caprioli17, hanusch19}.  The resulting \Qbars generally indicates T = 1 to 3 MK in the source regions of both impulsive and gradual events \citep{reames22a}, indicating normal coronal or active region plasma.  That result is somewhat surprising, in that one might expect strong heating and large populations of energetic electrons in reconnection current sheets, leading to elevated values of \Qbars in impulsive events.  High charge states in current sheets are observed in EUV images and UV spectra  \citep{reeves11, warren18, innes03, ciaravella08} and predicted by models \citep{shen13, shen23}.  The heating may be less intense in jets than in flares;  \citet{joshi20} found temperatures of 1.4-1.8 MK. 
However, \citet{paraschiv15} found that 75\% of the 16 jets they analyzed were between 1 MK and 3 MK, and 25\% were 4 MK and above.  Evidence for nonthermal electrons comes from type III radio bursts as well as X-rays \citep{paraschiv22, battaglia23, zhang23}. \citet{paterson23} found that NuSTAR observations of a jet were compatible with some nonthermal emission, but that the spectrum would have to be steep.

Here we examine the effects of non-Maxwellian electron distributions, on the initial temperatures derived from fitting $(A/Q)^\alpha$ to measured abundance distributions. In Section~2, we describe the calculations and results of the A/Q with the non-Maxwellian electron distributions and the temperature from the power-law fitting $(A/Q)^\alpha$ to a mean elemental abundance enhancement of impulsive SEP events. In Section~3, we discuss our results. In Section~4, we summarize our work and present conclusions. 

\section{Analysis and Results} \label{sec:analysis} 

We consider a plasma whose electron velocity distribution is out of thermal equilibrium as a result of particle energization in the solar corona in a magnetic reconnecting current sheet.  Acceleration processes may well yield electron and ion distributions that differ significantly from each other, but we seek an electron velocity distribution through which we can describe the ion charge states and their variations in $A/Q$. Classical particle systems reside at thermal equilibrium with their velocity distribution function stabilized into a Maxwellian distribution.  However, collisionless and correlated particle systems, such as the space and astrophysical plasmas, are characterized by a non-Maxwellian behavior, often described by the so-called kappa distribution, a more generalized form of particle velocity distribution where temperature and kappa are two independent parameters \citep{livadiotis13}.  Kappa distributions resemble a thermal core with a power-law tail.  They are general, physically meaningful, distribution functions that describe particle systems out of thermal equilibrium.  Use of empirical kappa distributions has become increasingly widespread across space and plasma physics, describing particles in the heliosphere, from the solar wind and planetary magnetospheres to the heliosheath \citep{livadiotis18}. We will use kappa functions here to model non-Maxwellian plasmas in the impulsive SEP acceleration region, which is likely to be a reconnecting current sheet.

\subsection{Mass to charge ratio with various kappa values} \label{subsec:kappa}

The source region temperatures for impulsive SEP events have been calculated using 
the dependence of the abundance enhancements in SEPs 
on the mass-to-charge ratios $(A/Q)$ \citep{reames14a, reames16, reames18a}. 
Elements with low first ionization potential (FIP$<$10 eV) are enhanced
in the corona by about a few factors with respect to the 
photosphere \citep[e.g.][]{feldman92, schmelz12}. 
The abundance enhancements observed in SEPs show the ``FIP effect" \citep{schmelz12,reames18b}, 
but the patterns of the enhancements differ in impulsive and gradual SEPs \citep[e.g.][]{reames20}. 
The heavy elements are more enhanced in the impulsive SEPs, while the gradual events are similar to 
the corresponding element abundances in the solar corona \citep{reames14a}. 

The temperatures of the impulsive SEPs' source region are found to be about 2$\sim$ 3 MK by fitting the SEP abundance power-law distribution vs $(A/Q)^\alpha$, which depends on the electron temperature \citep[e.g.][]{reames20}. These are consistent with recent studies based on EUV observations of the active region that produced an impulsive SEP event \citep{bucik21}, though the active region is far larger than the SEP source region.  However, these works assume a Maxwellian electron distribution in calculating the average charge state, \Qbars,  while the electrons are likely to be out of thermal equilibrium during the energization.  

We examine how much the temperature estimation from the A/Q would differ with various kappa values in a kappa function representing high-energy tails deviating from a Maxwellian velocity distribution. We use the KAPPA package \citep{dzifcakova21}  recently upgraded with the CHIANTI 10 atomic database \citep{delzanna21} with the $\kappa$ distributions defined as in \citet{dzifcakova21}, 

\begin{equation} \label{eq:1}
f_k(E)dE = A_k \frac {2} {\sqrt{\pi}(k_B T)^{3/2}} \frac {E^{1/2}dE} { (1 + {\frac{E} {(\kappa-3/2)k_BT} )^{\kappa + 1}} }, 
\end{equation}

\noindent
where E is the kinetic energy,  $k_B$ is the Boltzmann constant, T is the temperature, $A_k=\Gamma(\kappa+1)/\Gamma (\kappa -1/2)/(\kappa-3/2)^{3/2}$ is a normalization constant (see details in \citealp{dzifcakova21}). We calculate the averaged charge states $\langle Q \rangle_{T}$ at temperature T for each element (atomic number  Z) and the $\kappa$ values ($\kappa$),
\begin{equation} \label{eq:2}
\langle  Q  \rangle_{T} =  \sum_{i=1}^{Z-1} {i} {q{(i,T)}}, 
\end{equation}

\noindent
where $q(i,T)$ is the ion fraction of charge state $i$  of each $\it{Z}$ and $\kappa$ as a function of the temperature, obtained from the KAPPA package. 
The ion fractions depend on the density and temperature in time-dependent nonequilibrium ionization. In this calculation, we assume ionization equilibrium since the particle acceleration timescales can be much shorter than the ionization/recombination timescales. 

\begin{figure}
\centering
\includegraphics[width=100mm]{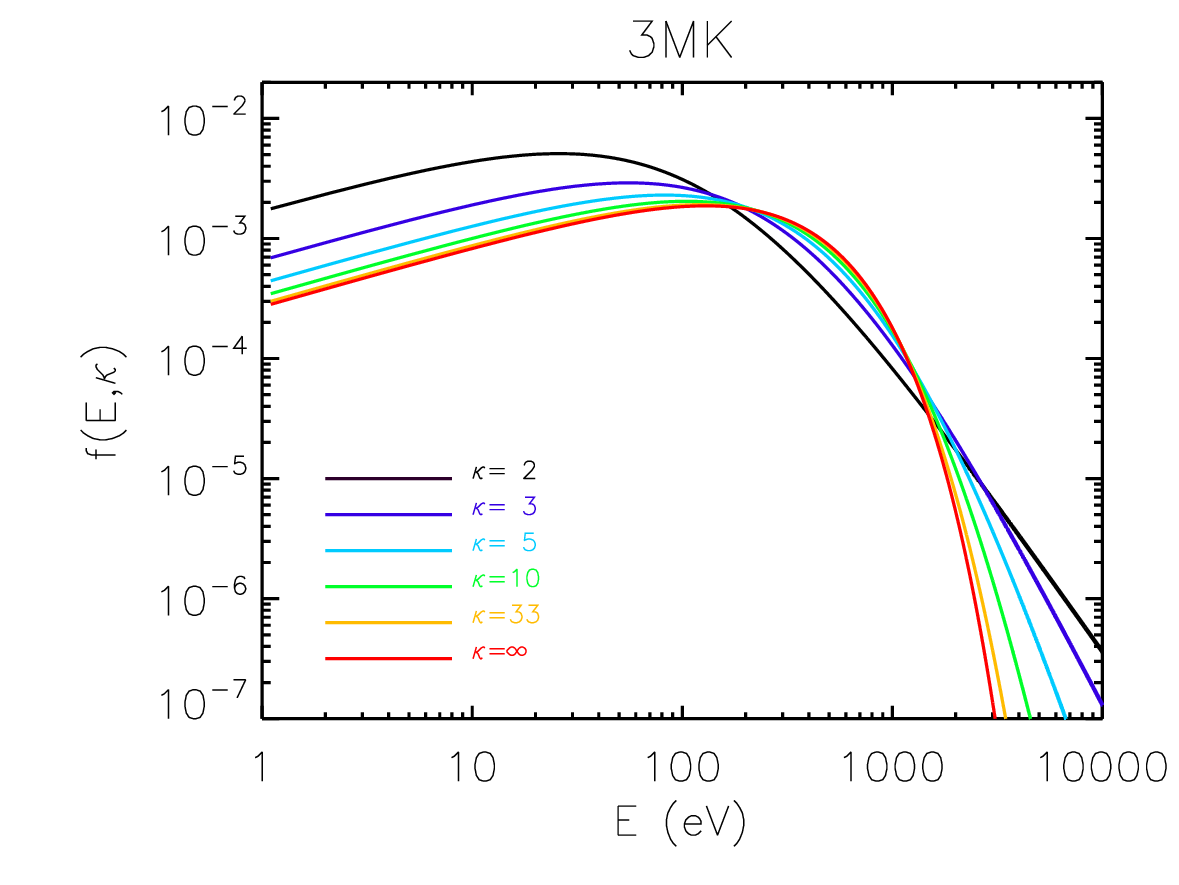}
\caption{$\kappa$-distributions with various $\kappa$ values at 3 MK} 
\label{fig:ekappa}
\end{figure}

Figure~\ref{fig:ekappa} shows the electron kinetic energy distributions at 3~MK with various $\kappa$ values calculated using the KAPPA package.  When the $\kappa$ value goes to infinity, the distributions are the same as the Maxwellian distribution 
(see equation~\ref{eq:1}), and the lower $\kappa$ value means the higher deviation from the Maxwellian distribution. 
At an extreme case of $\kappa$=2, the energy distribution is higher at lower energy (a few to several tens of eV) and represents the high-energy tails at the higher than a few thousand eV. 

\begin{figure}
\centering
\includegraphics[width=80mm]{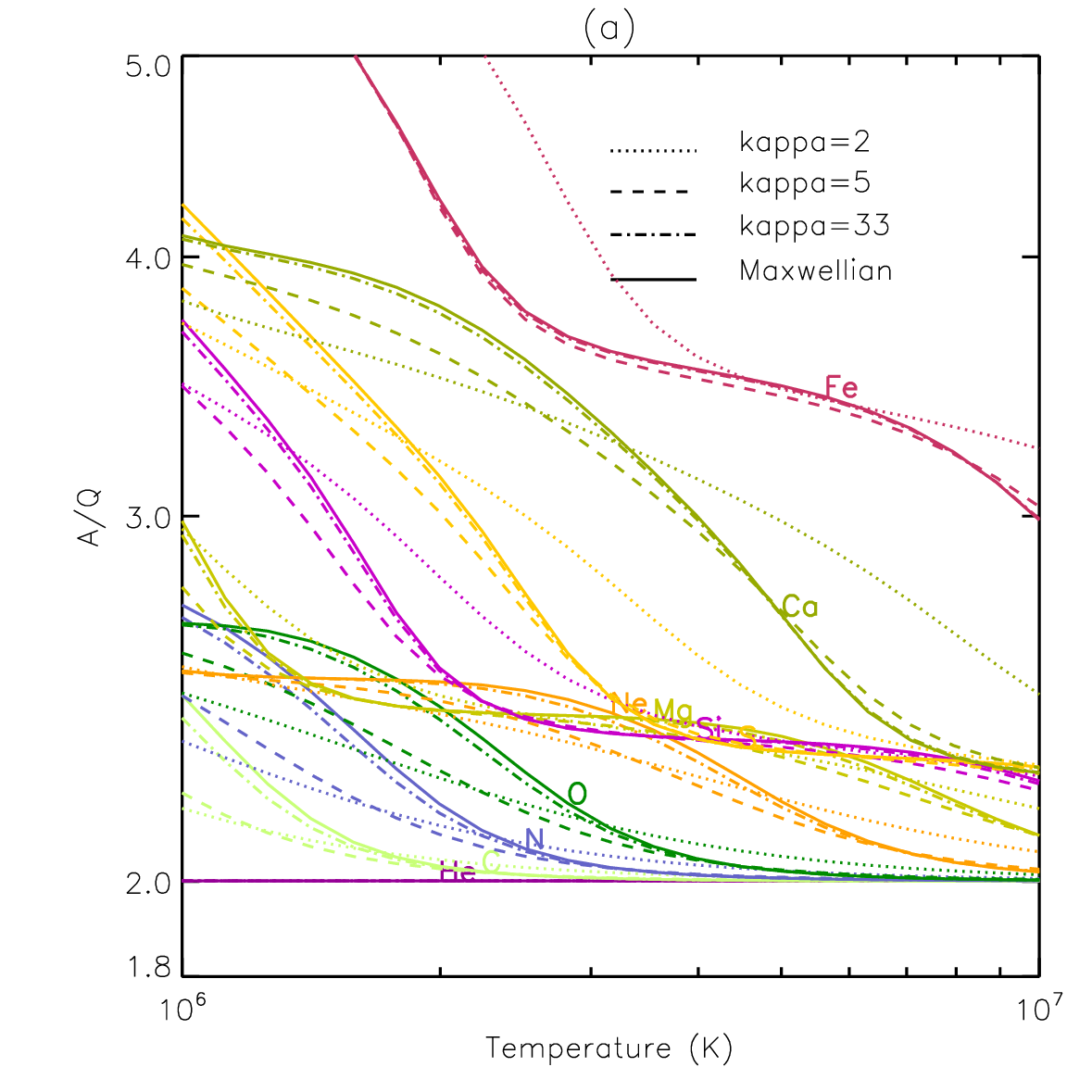}
\includegraphics[width=80mm]{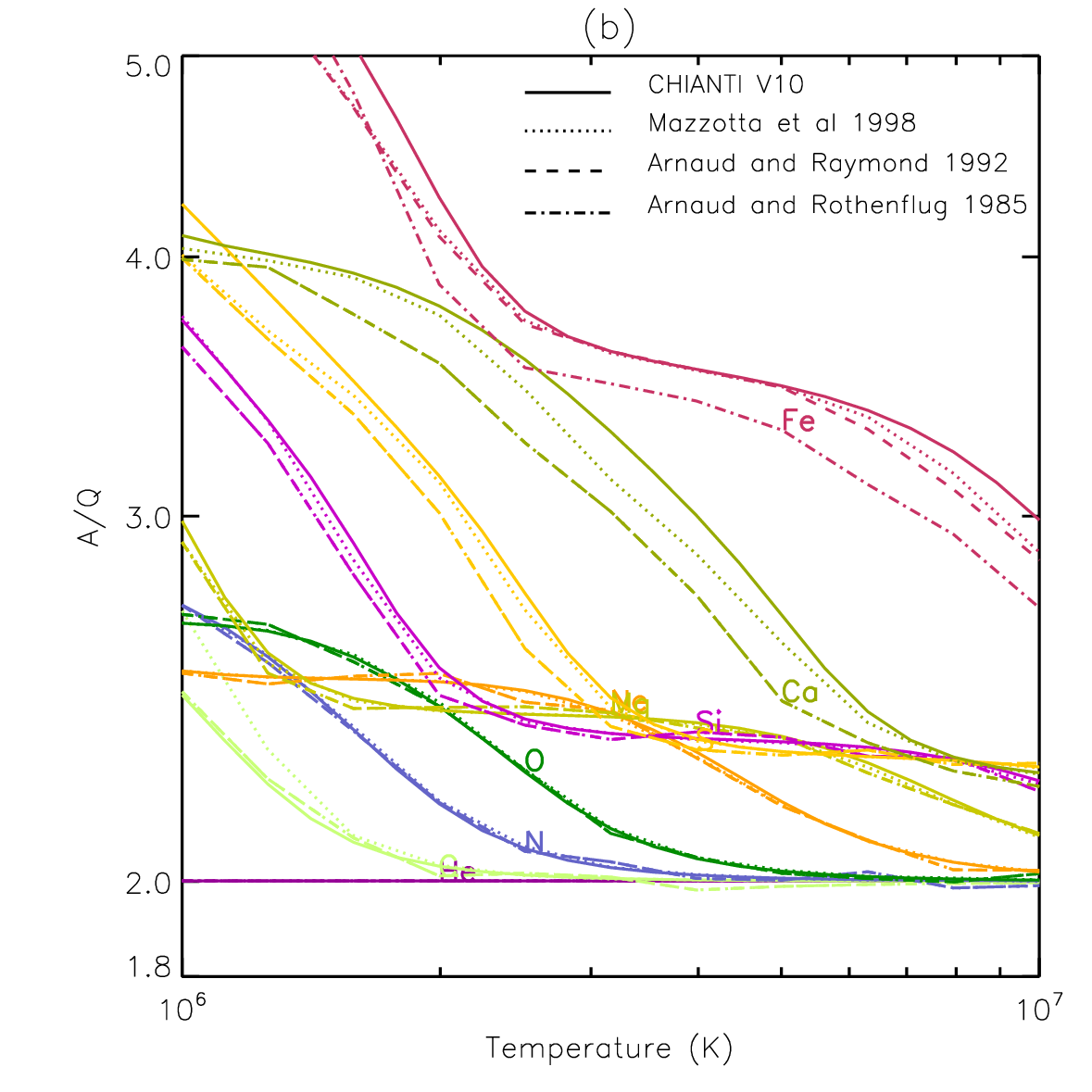}
\caption{(a) Mass to charge ratios from ionization equilibrium with various $\kappa$ values (b)Mass to charge ratios from ionization equilibrium with the Maxwellian velocity distribution, from various atomic rate compilations: CHIANTI V10 \citep{delzanna21}, Mazzotta et al. 1998, Arnaud and Raymond 1992, and Arnaud and Rothenflug 1985.}
\label{fig:Taq}
\end{figure}

\begin{figure}
\centering
\includegraphics[width=50mm]{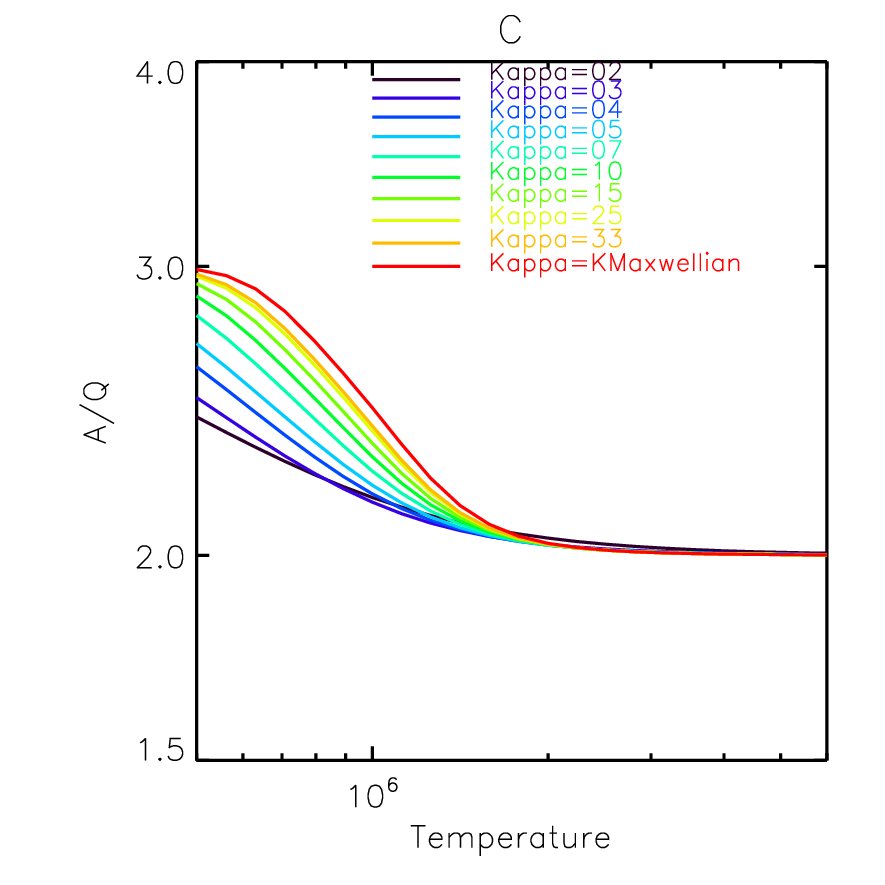}
\includegraphics[width=50mm]{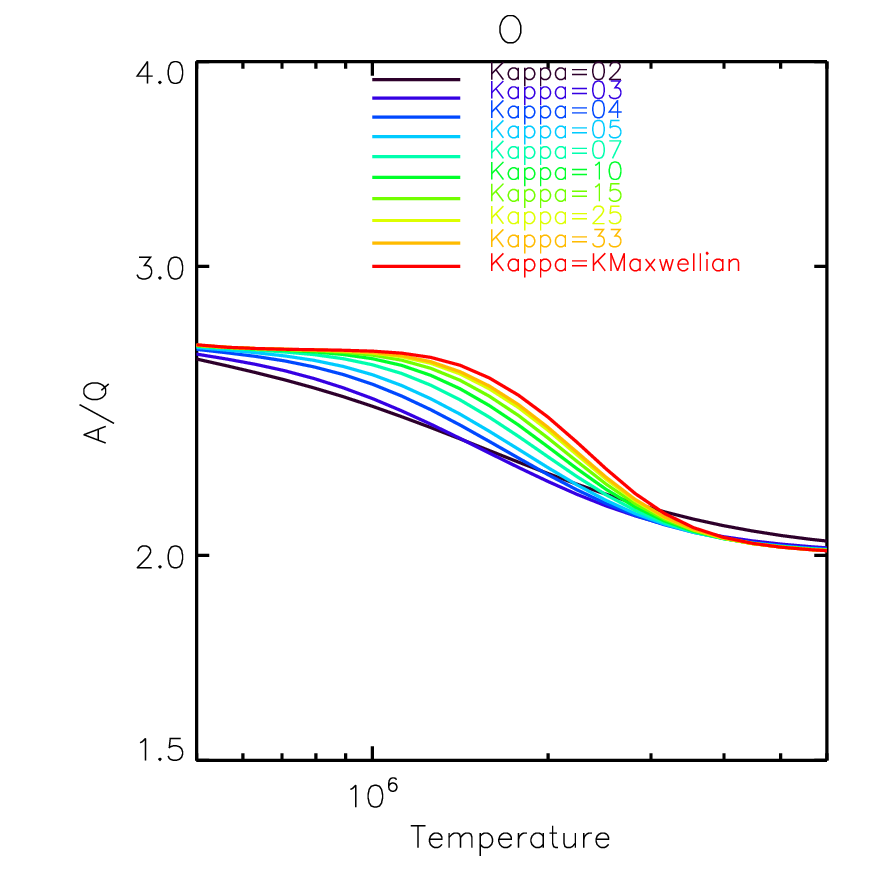}
\includegraphics[width=50mm]{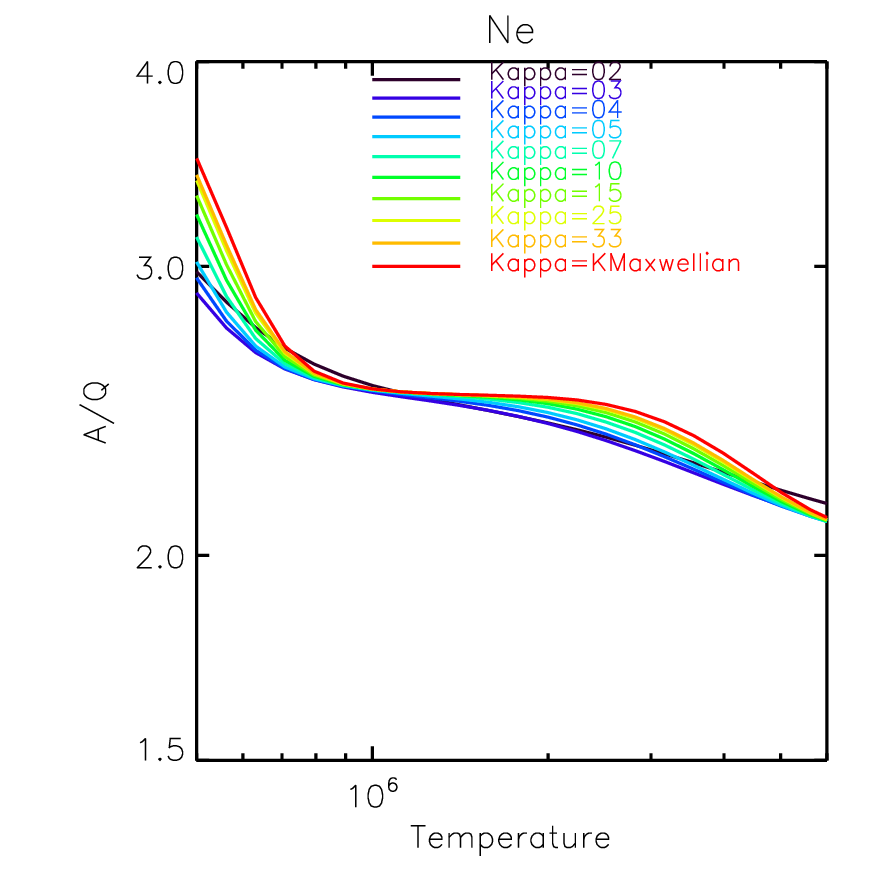} \\
\includegraphics[width=50mm]{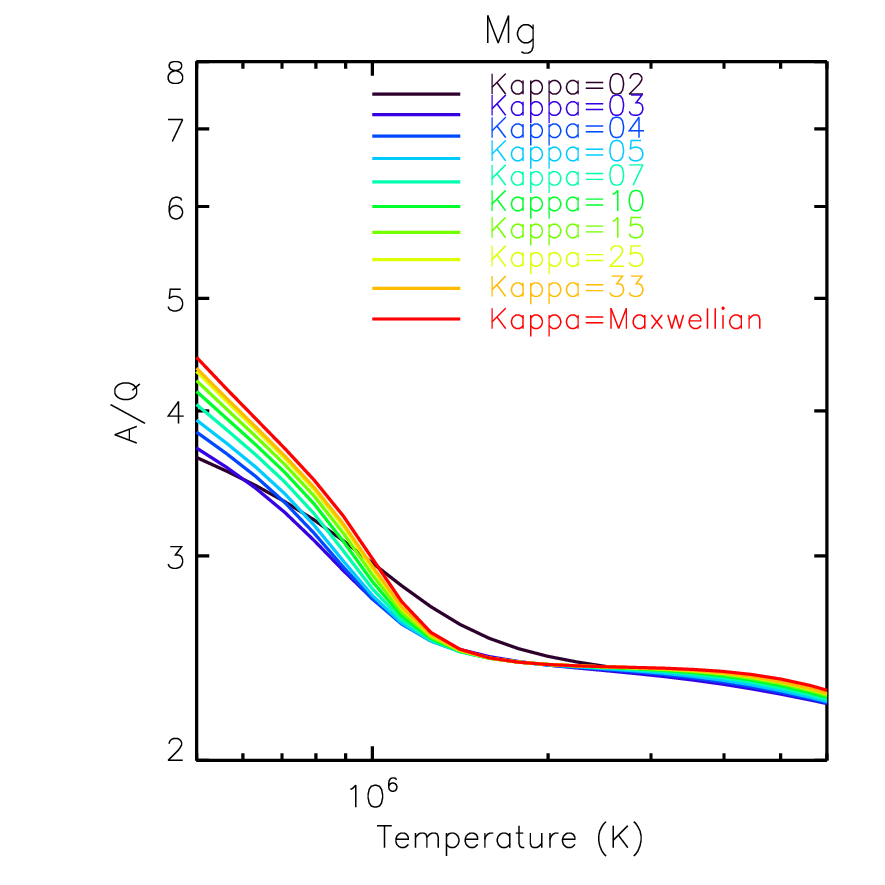}
\includegraphics[width=50mm]{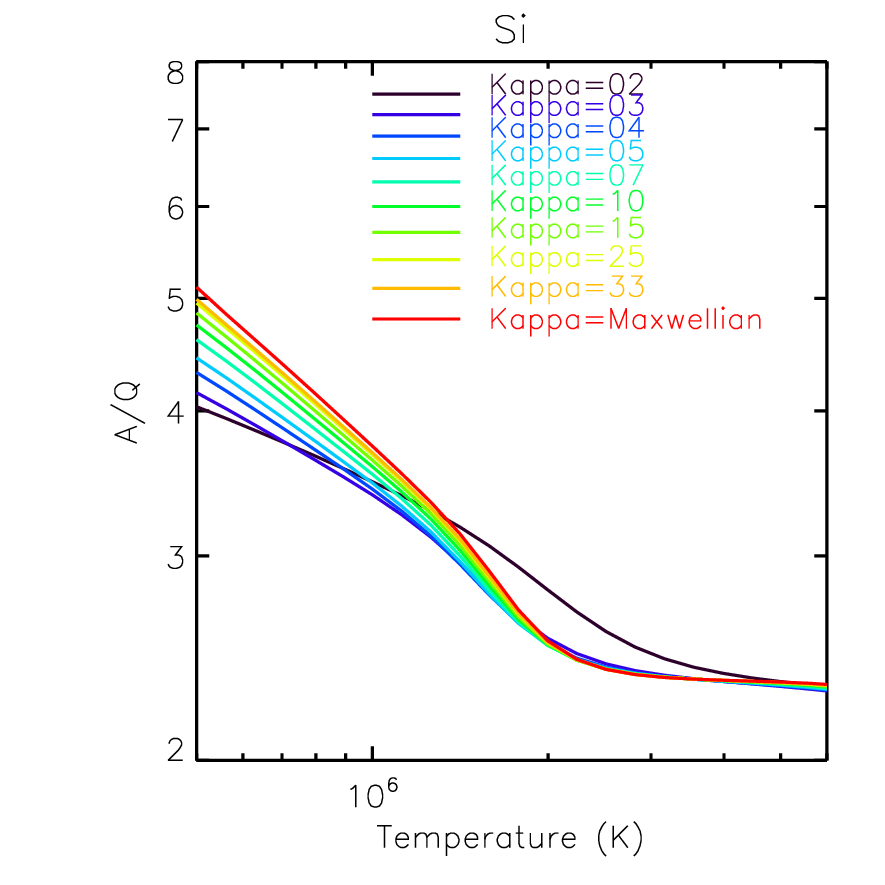}
\includegraphics[width=50mm]{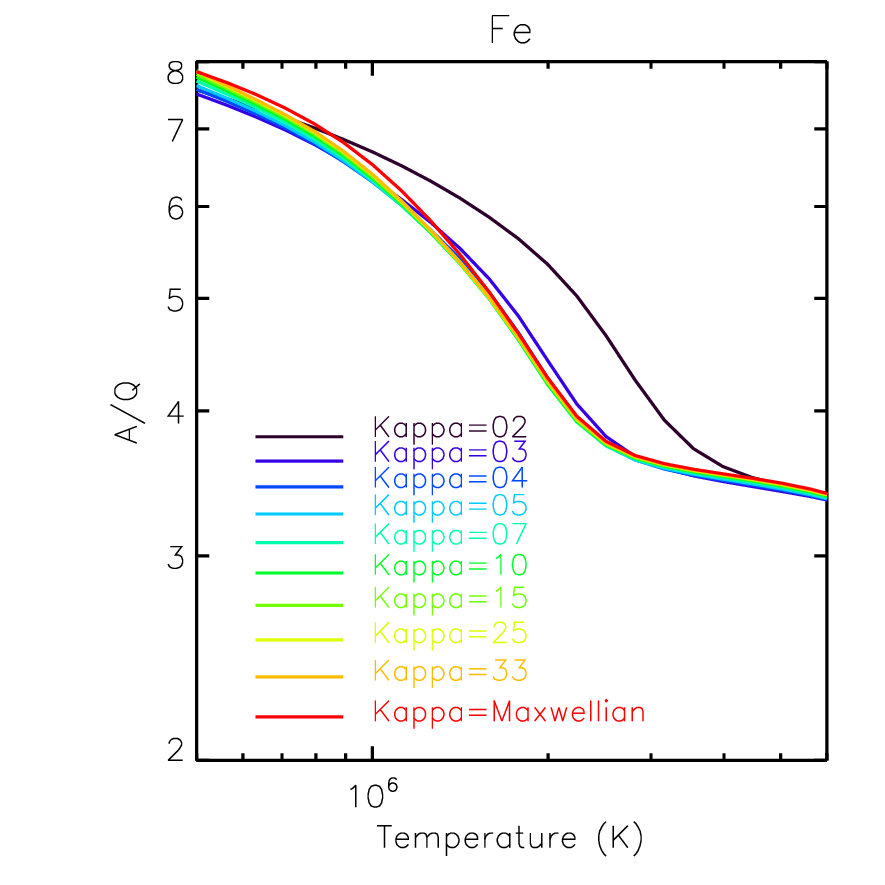}
\caption{Mass to charge ratios in ionization equilibrium with various $\kappa$ values for C, O, Ne, Mg, Si, and Fe elements}
\label{fig:aqele}
\end{figure}

\begin{figure}
\centering
\includegraphics[width=80mm]{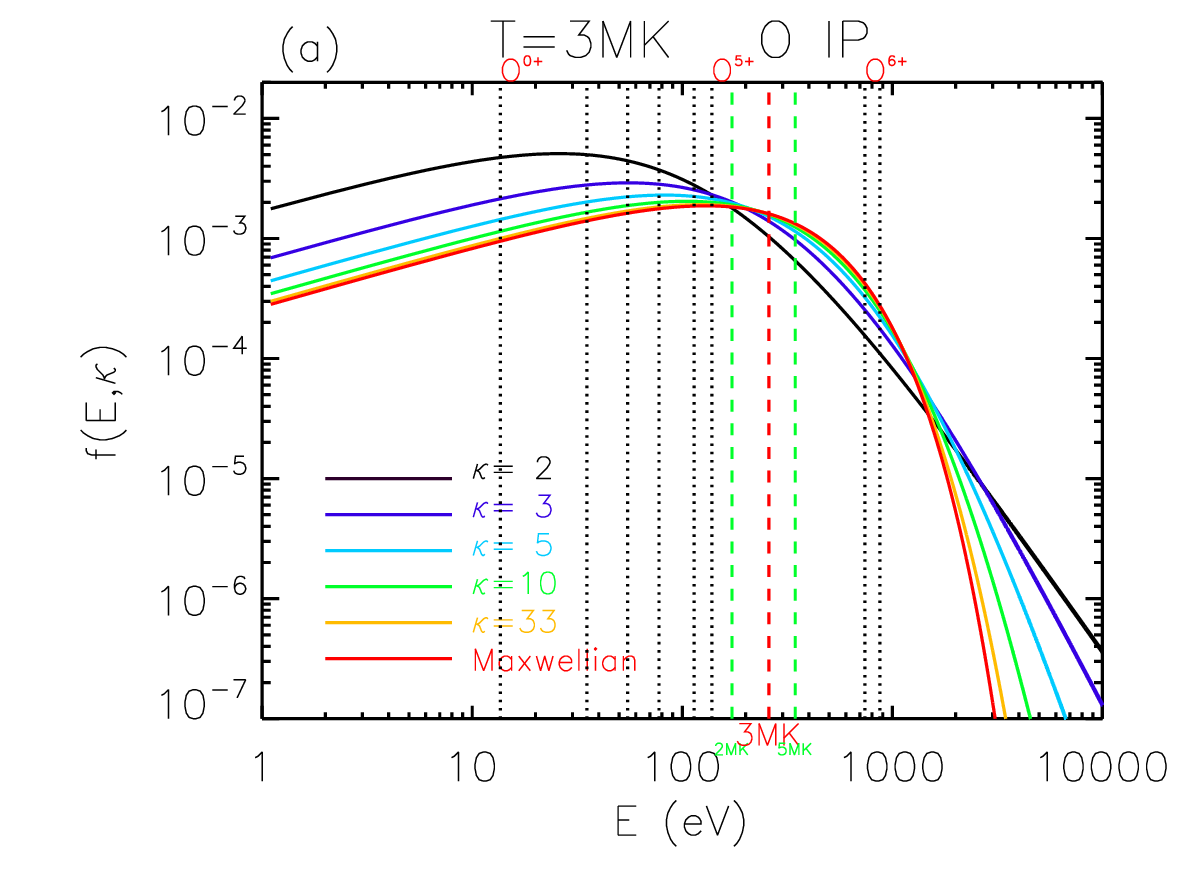}
\includegraphics[width=80mm]{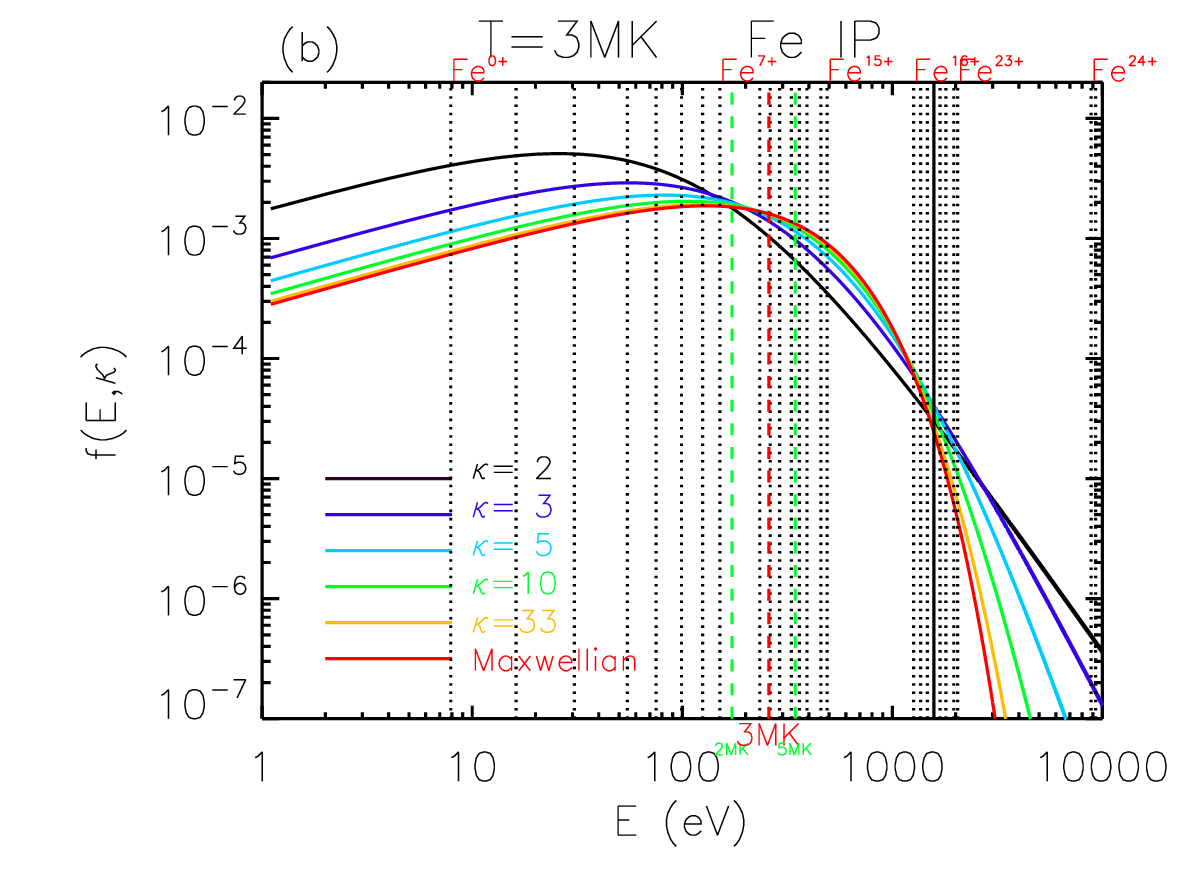}
\caption{$\kappa$-distributions with various $\kappa$ values at 3 MK overplotted with ionization potentials of O and Fe ions. 
The energy distributions with various $\kappa$ values are the same as in Figure~\ref{fig:ekappa}. 
We added vertical lines of the ionization potentials and coronal temperatures in eV. 
Black dotted vertical lines represent the ionization potentials of each ion from 0+ (lowest energy) to +Z-1 (highest energy).
The red (3~MK) and green (2 and 5~MK) dashed lines represent the usual coronal temperatures in active regions 
that roughly point to the dominant coronal ionization stages around these temperatures.}
\label{fig:ekappaip}
\end{figure}

Figure~\ref{fig:Taq}~(a) shows the $A/Q$ (where $Q$ is the average charge state as in Eq.(2)) calculated using the KAPPA package with kappa distributions of $\kappa$=2, 5, 33, and Maxwellian distribution.  
The differences in $A/Q$ to that from the Maxwellian distribution are generally larger with smaller $\kappa$ values. 
We show $A/Q$ calculated for the elements, C, O, Ne, Mg, Si, and Fe with more $\kappa$ values in Figure~\ref{fig:aqele}. 
The differences from different $\kappa$ values are relatively larger at lower temperatures and show the general effect of enhanced ionization (thus lower A/Q) with lower kappa values such as C, O, and Ne in upper panels in Figure~\ref{fig:aqele}. 
This is illustrated in Figure~\ref{fig:ekappaip}~(a) for O ions. At lower temperatures, the peak of the kappa distribution shifts toward lower energy, thus the number of electrons above, e.g., the ionization potential of O$^{+6}$ (740 eV) increases for lower $\kappa$ because the power-law tail begins to dominate. This leads to the effect of $< Q_O >$ increasing (A/Q decreasing) with decreasing $\kappa$ for temperatures below about 2 MK (Figure~\ref{fig:aqele}).
In the case of $\kappa$=2 for the higher Z elements, however, this trend is reversed at temperatures around a few MK, especially in Fe (lower panels in Figure~\ref{fig:aqele}). 
This occurs both because there are fewer electrons at 250-350 eV energies 
that can ionize the Si and Fe ions typically found at coronal temperatures, 
and because there are more electrons at energies of a few tens of eV 
that can drive dielectronic recombination (Figure~\ref{fig:ekappaip}~(b)). 

\begin{figure}
\centering
\includegraphics[width=80mm]{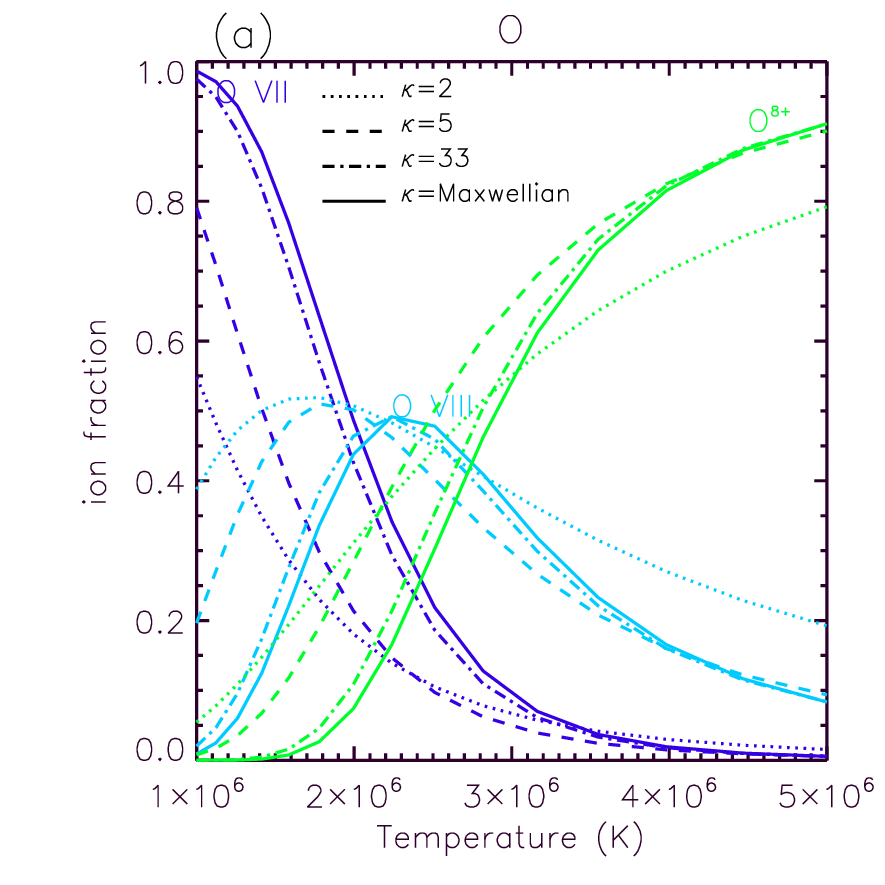}
\includegraphics[width=80mm]{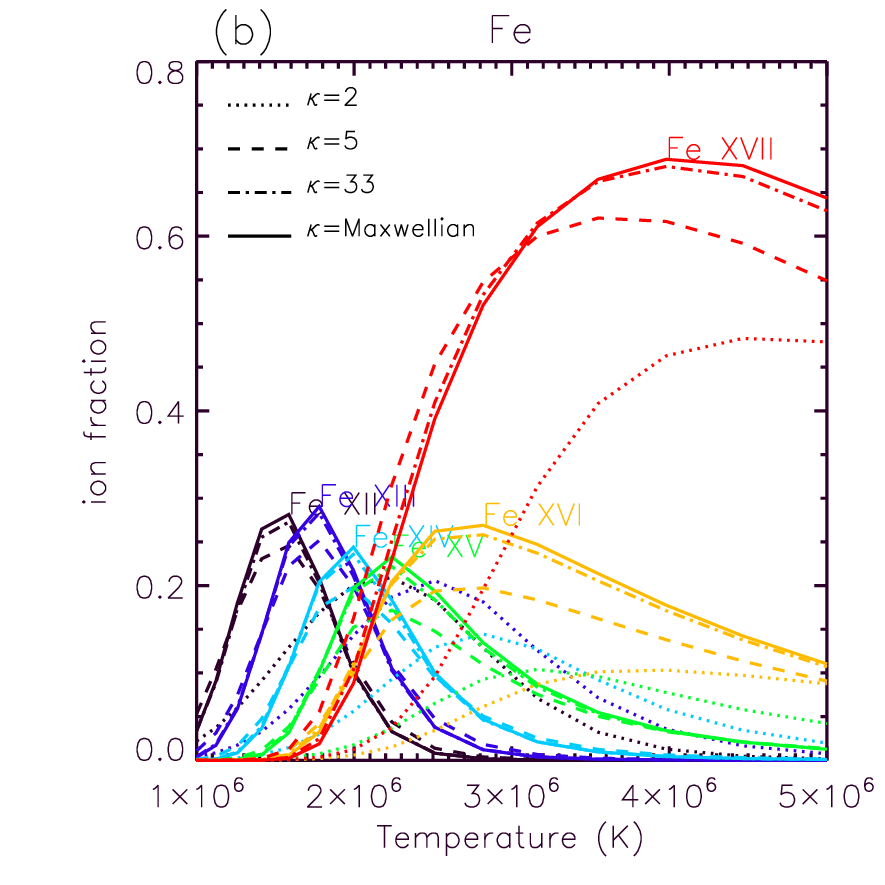}
\caption{(a) O~VII, O~VIII,  and fully stripped O$^{8+}$ ion fractions as a function of temperature, with $\kappa$=2, 5, 33, and Maxwellian distributions. 
(b) Fe~XII, Fe~XIII, Fe~XIV, Fe~XV, Fe~XVI, and Fe~XVII ion fractions with $\kappa$=2, 5, 33, and Maxwellian distributions.}
\label{fig:ionf}
\end{figure}

Figure~\ref{fig:ionf} shows several ion fractions of O and Fe, as a function of temperature, representing low and high Z elements, respectively. 
In the case of O, the ion fractions with the lower $\kappa$ are greater at lower temperatures (Figure~\ref{fig:ionf} (a)). 
In other words, the peaks of the ion fractions with lower $\kappa$ are left shifted to the lower temperatures. 
It indicates that the charge states are higher at the coronal temperature, so A/Q decreases. 
On the other hand, in the case of Fe, the ion fractions with $\kappa=2$ shift toward higher temperatures (Figure~\ref{fig:ionf} (b)), 
leading to lower charge states at the coronal temperature with lower $\kappa$, so A/Q increases. 
However, we find that the differences in the A/Q between a Maxwellian and an extreme kappa distribution are only about 10-30\%.

Earlier publications used different atomic data compilations \citep[e.g.][used \citet{arnaud85} for elements below Fe, and \citet{arnaud92} for Fe]{reames14b}. 
To investigate the impact from atomic data, we also compare A/Q in ionization equilibrium with the Maxwellian distribution from CHIANTI V10, \citet{mazzotta98}, \citet{arnaud85}, \citet{arnaud92} in Figure~\ref{fig:Taq}~(b).  We find that the differences in the $A/Q$ among them are small (at most 30\% in high-Z elements of Ca and Fe).  In addition, we find that the $A/Q$ approaches the value of ``two" at higher temperatures for all elements (except Fe) since these elements are almost fully ionized at higher temperatures of about 10~MK.  Therefore, the $A/Q$ would not be an appropriate parameter to estimate the high temperature plasma at the source region. 

\subsection{Power law fitting of mass to charge ratio vs. abundance enhancements} \label{subsec:plaw}

 A comprehensive study of 111 impulsive events
observed with the Low Energy Matrix Telescope (LEMT) on the Wind spacecraft was carried
out by \citet{reames14b, reames14a, reames15}, selected with the requirement of (Fe/O)/0.131$>$4 where
0.131 is the assumed coronal value of Fe/O. CMEs of median widths (75\textdegree) and speeds (600
km s$^{-1}$) were associated with 69\% (66 of 96) of the events with available LASCO coronagraph
observations \citep{reames14b}.

Assuming $A/Q$ values characteristic of a coronal plasma temperature range of 2.5-3
MK, \citet{reames14b} plotted the mean elemental enhancements from He to Pb versus their
A/Q for 111 impulsive events. They found the power-law least-squares best fit slope of 3.64$\pm$0.15 
(see Figure~8 in their paper). In their subsequent work, \citet{reames14a} introduced a least squares
best fit methodology to determine both the most probable power of abundance
enhancement versus $A/Q$ and the best value of source plasma temperature T to determine Q.
They applied the technique to each of the 111 impulsive SEP events, then compared the results
to properties of associated CMEs, flares, and $^3$He-rich events. 

\begin{figure}
\centering
\includegraphics[width=80mm]{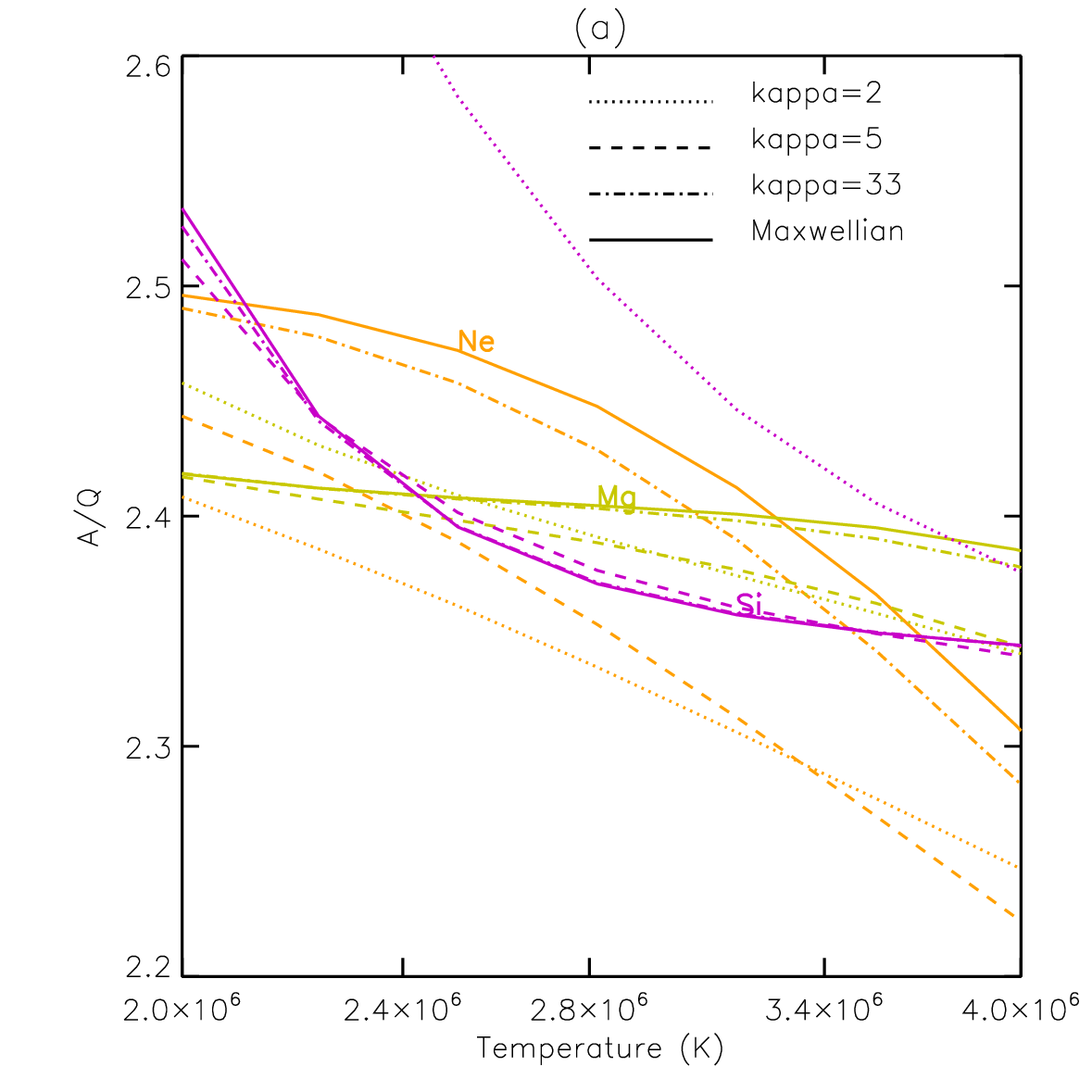}
\includegraphics[width=80mm]{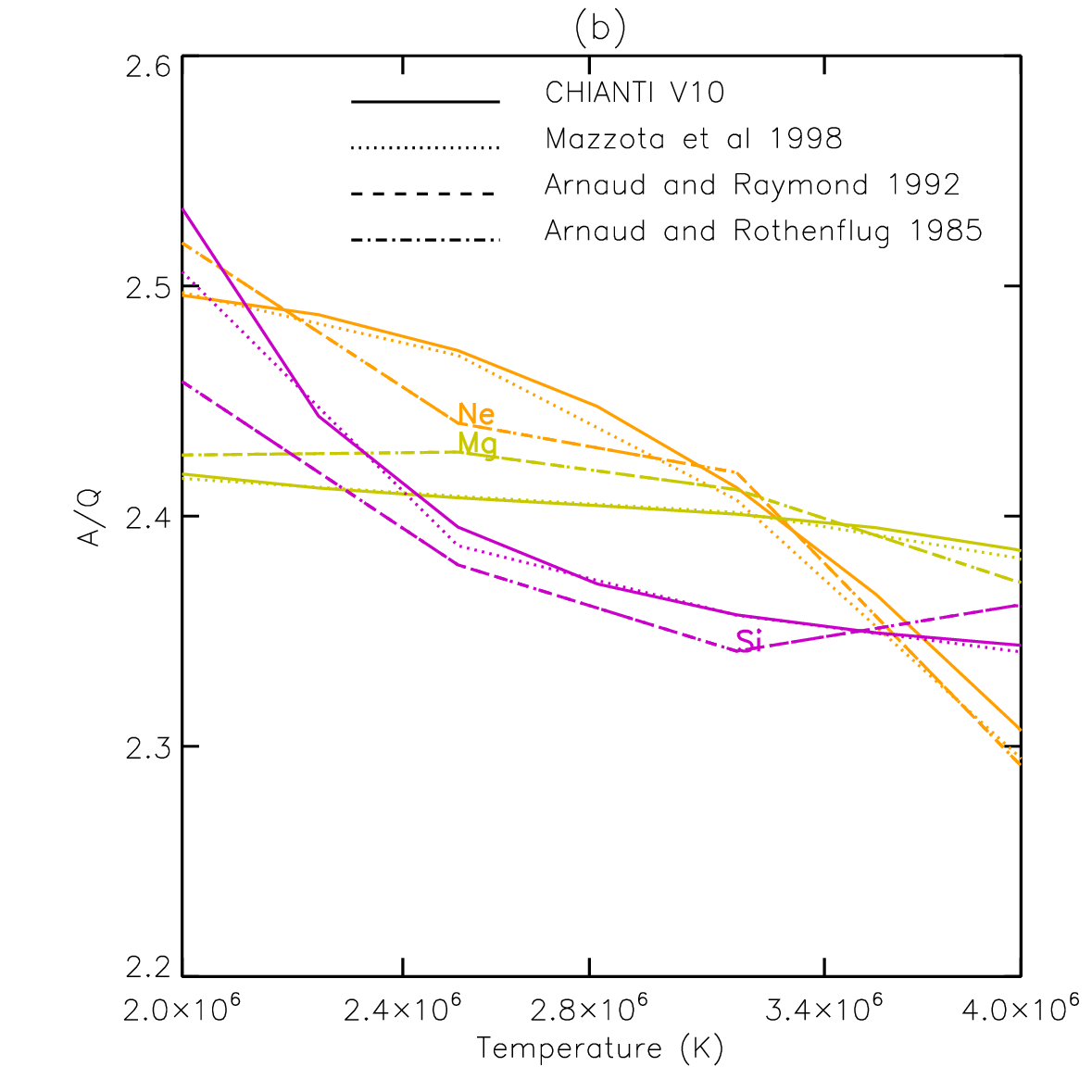}
\caption{Same as in Figure~\ref{fig:Taq} for the elements, Ne, Mg, and Si, zoomed in to a narrow temperature range}
\label{fig:Taqss}
\end{figure}

\begin{figure}
\centering
\includegraphics[width=100mm]{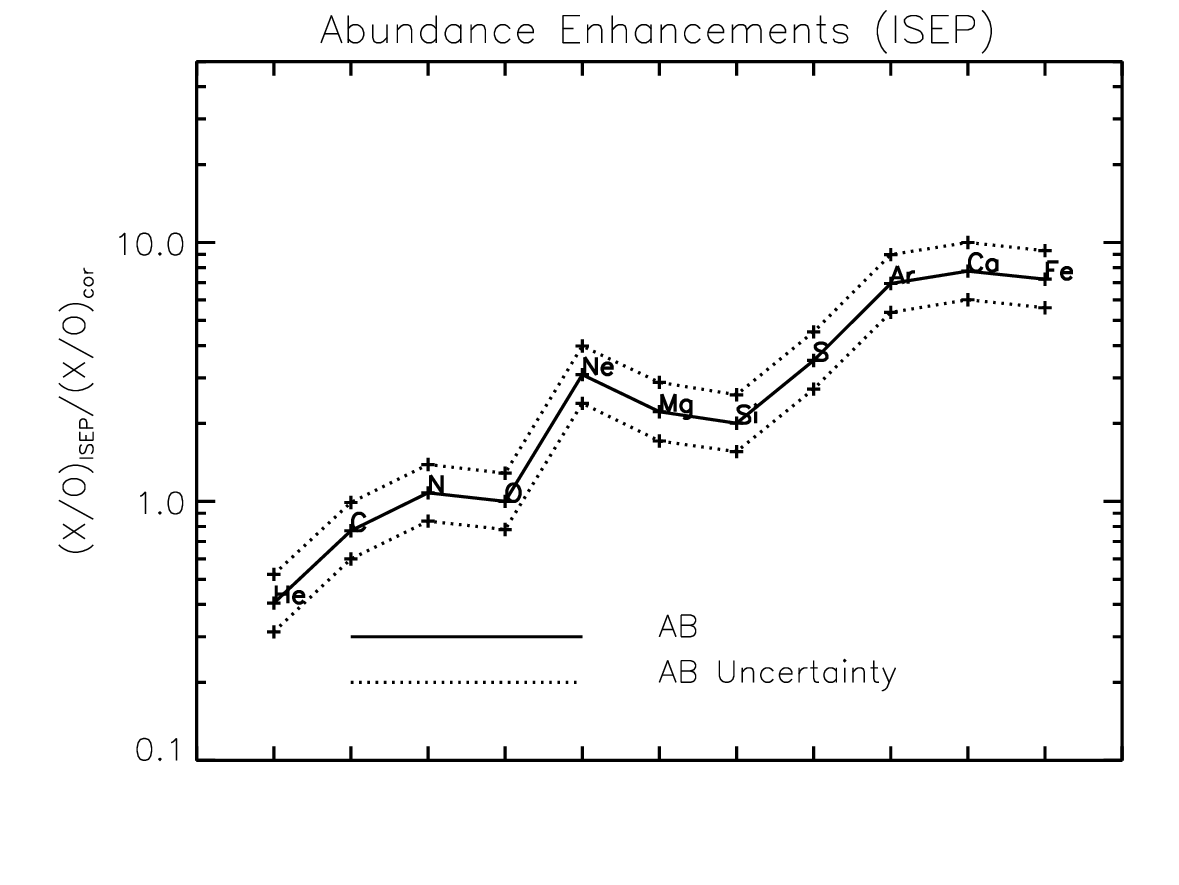}
\caption{The mean abundance enhancements (AB) in the impulsive SEP (ISEP) events in \citep{reames14b}.
The uncertainties are calculated using the observed uncertainties in \citet{reames14b} and 
assuming 25\% uncertainty in the coronal abundance} 
\label{fig:aben}
\end{figure}

The $A/Q$ of most elements tends to follow the order of atomic number Z (Figure~\ref{fig:Taq}). 
But, around 2$\sim$3~MK, the $A/Q$ is opposite as $(A/Q)_{Ne}$ $>$ $(A/Q)_{Mg}$ $>$ $(A/Q)_{Si}$ in Figure~\ref{fig:Taqss}.
The observed enhancements in impulsive SEP events also follow the order Ne $>$ Mg $>$ Si (Figure~\ref{fig:aben}). 
It has been discussed that this feature accounts for the best fit at 2$-$3~MK in the power-law fit of the abundance enhancements of the impulsive SEP events\citep{reames14b, reames18a, reames21}.
These features are applicable in Maxwellian plasma and larger values of $\kappa$.  In the extreme case of $\kappa$=2, $A/Q$ follows the order of Z, $(A/Q)_{Si}$ $>$ $(A/Q)_{Mg}$ $>$ $(A/Q)_{Ne}$ (Figure~\ref{fig:Taqss}~(a)). The$A/Q$ of Maxwellian plasma in ionization equilibria from various atomic rate compilations keep the opposite order (Figure~\ref{fig:Taqss} (b)). 
Thus, we examine these effects on the power-law fit for estimating the source region temperature. 

We adopt the mean abundance enhancement in impulsive SEP events explained above in \citet{reames14b} (see Table~1 and Figure~8 in their paper).
The abundance enhancement is expressed as the ratio of the observed mean element abundances (X) divided by Oxygen (O) $((X/O)_{ISEP}$) over 
the enhancements relative to a coronal abundance based on the \citet{feldman92} coronal abundance set
(Sun$\_$Coronal$\_$1992$\_$feldman$\_$ext.abund in CHIANTI) ($(X/O)_{cor}$). 
Note that \citet{reames14b} use the enhancements relative to mean element abundances of gradual SEP events (Table 1 in their paper). 
It has been known that the abundances in gradual events are closely related to the coronal abundances \citep{reames14a}.
We present the abundances in Table~\ref{tab:abundance}.   Figure~\ref{fig:aben} shows that the mean abundance enhancements of the elements mostly follow the order of Z, except the opposite order of Ne $>$ Mg $>$ Si. 

\begin{figure}
\centering
\includegraphics[width=50mm]{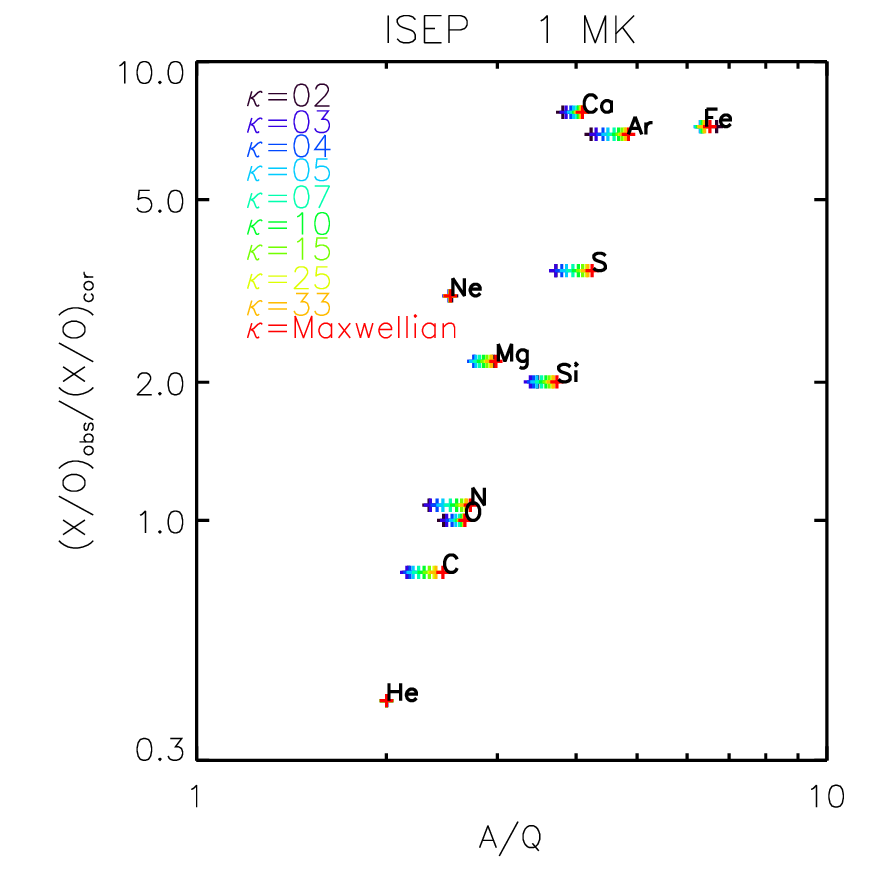}
\includegraphics[width=50mm]{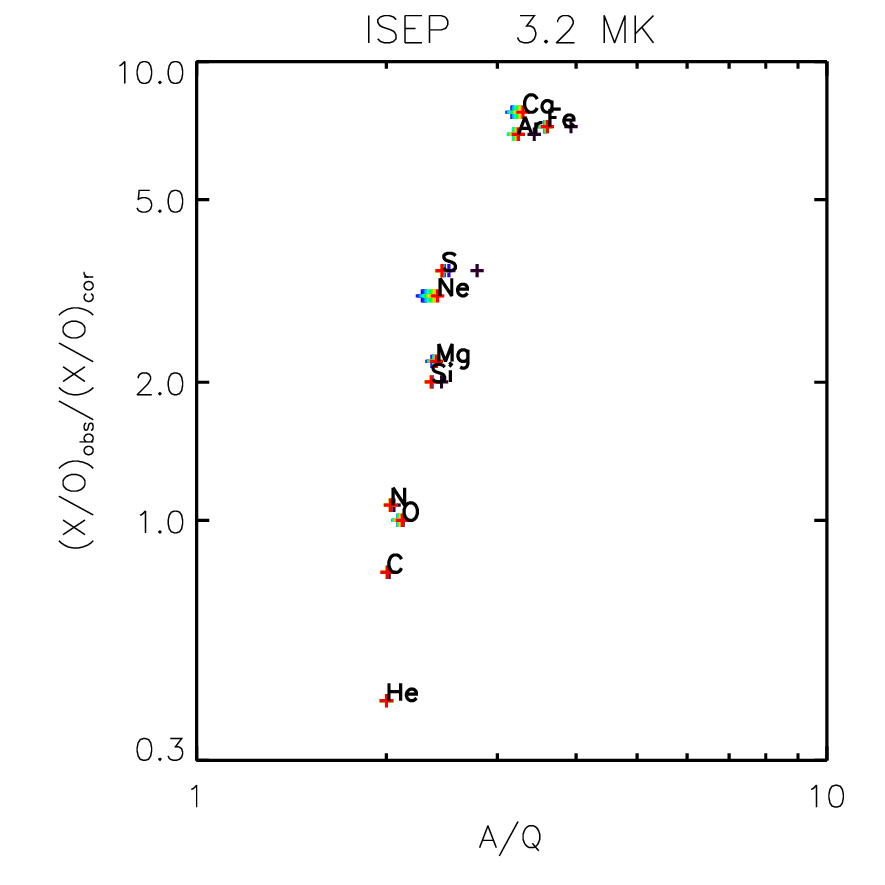}
\includegraphics[width=50mm]{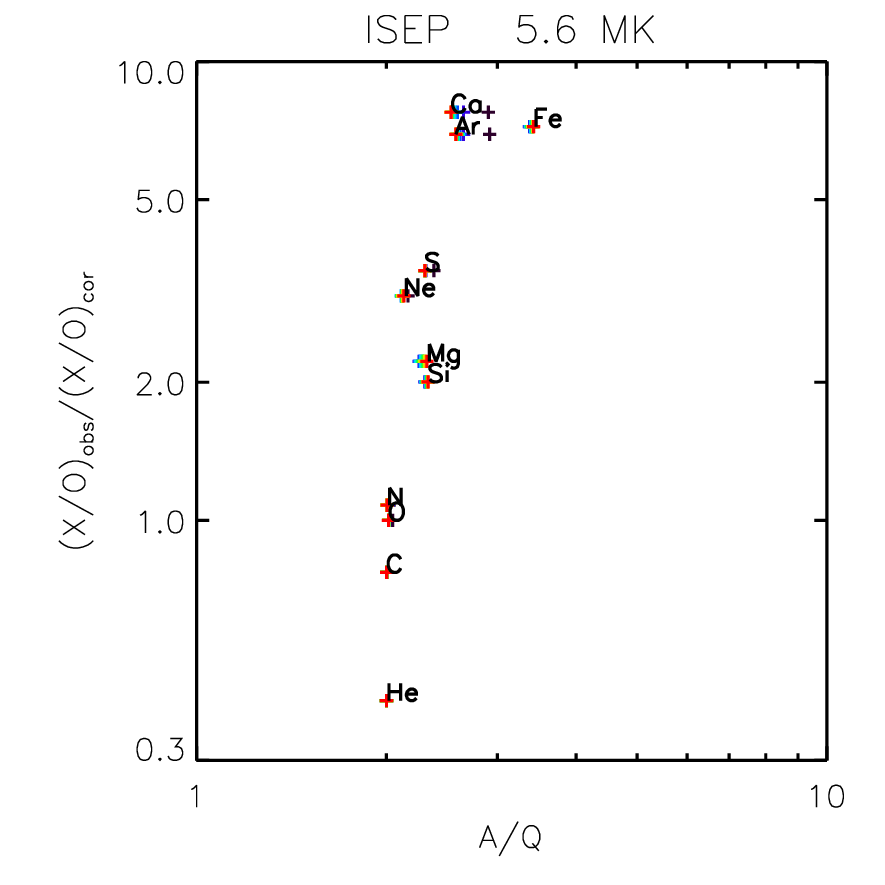}
\caption{Abundance enhancements versus $A/Q$ 
at the temperature of  1 MK, 3.2 MK, and 5.6 MK  with various $\kappa$ values. 
The color codes are the same as in Figure~\ref{fig:aqele}.}
\label{fig:abenaq}
\end{figure}

Figure~\ref{fig:abenaq} shows the abundance enhancement with $A/Q$ at 1, 3.2, and 5.6~MK. 
As discussed in Section~\ref{subsec:kappa}, the differences of $A/Q$ with various $\kappa$ values at higher temperatures are modest,  
so the abundance enhancements with $\kappa$ values are most spread out at lower temperatures around 1 MK. 

The strong power-law dependence of the abundance enhancements on A/Q in impulsive SEP events has been 
explained by the high $A/Q$ particles gaining energy more easily than
protons and enhancing their abundances \citep{drake09}. 
The power-law fits yield estimated source region temperatures of impulsive SEP events  
to be $\sim$2-5~MK using the charge states calculated assuming Maxwellian distribution \citep{reames14a,reames14b}. 
We test the power-law enhancement of element abundances as a function $A/Q$ with various $\kappa$ values 
by using a relation between the abundance enhancement (AB) and $A/Q$, 

\begin{equation}
AB \propto (A/Q)^{\alpha} 
\end{equation}

\noindent
Applying a decimal logarithm, we fit a first order linear equation, 

\begin{equation}
Log_{10}(AB) = \alpha Log_{10} ( \frac {A} {Q} ) + b, 
\end{equation}

\noindent
where $\alpha$ is a slope representing the power-law index and b is a Y-intercept. 
Then, we find a minimum reduced $\chi_r^2=\chi^2/(n-2)$, 

\begin{equation}
\chi^2 = \sum_{Z=1}^{N} (\frac {y_Z - b - \alpha x_Z} {\sigma_Z})^2, 
\end{equation}

\noindent
where $y_Z$ is $Log_{10}(AB)$, $x_Z$ is $Log_{10}(A/Q)$, 
$\sigma_{Z}$ is the uncertainty calculated using the observed uncertainties in \citet{reames14b} 
and assuming 25\% uncertainty in the coronal abundance, 
and n is the number of elements. 

We use 10 elements (C, N, O, Ne, Mg, Si, S, Ar, Ca, and Fe), not including He and the higher Z elements (Z=34$-$82) from \citet{reames14b}.
In previous studies, \citet{reames17} found that the He/O ratio varies widely from event to event, 
and it causes a second minima at high temperatures in the fits of gradual events. 
Later, \citet{reames22b} finds that the shock-driven SEP events show more He (see Figure 13 in the paper). 
In addition, sounding rocket observations reveal that helium is depleted in the equatorial regions during the quiet Sun \citep{moses20}. 
For the reason of this, we exclude He for the power-law fitting.
We show the fitting parameters at the second minimum in Table~\ref{tab:pwdi}.

\begin{figure}
\centering
\includegraphics[width=80mm]{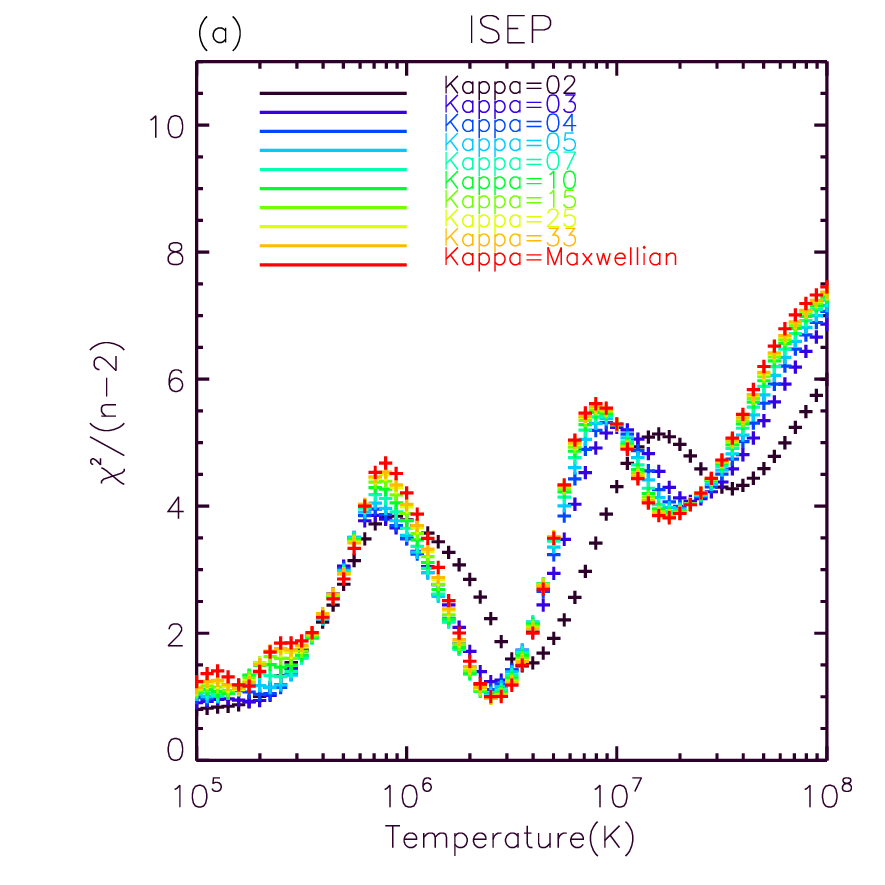}
\includegraphics[width=80mm]{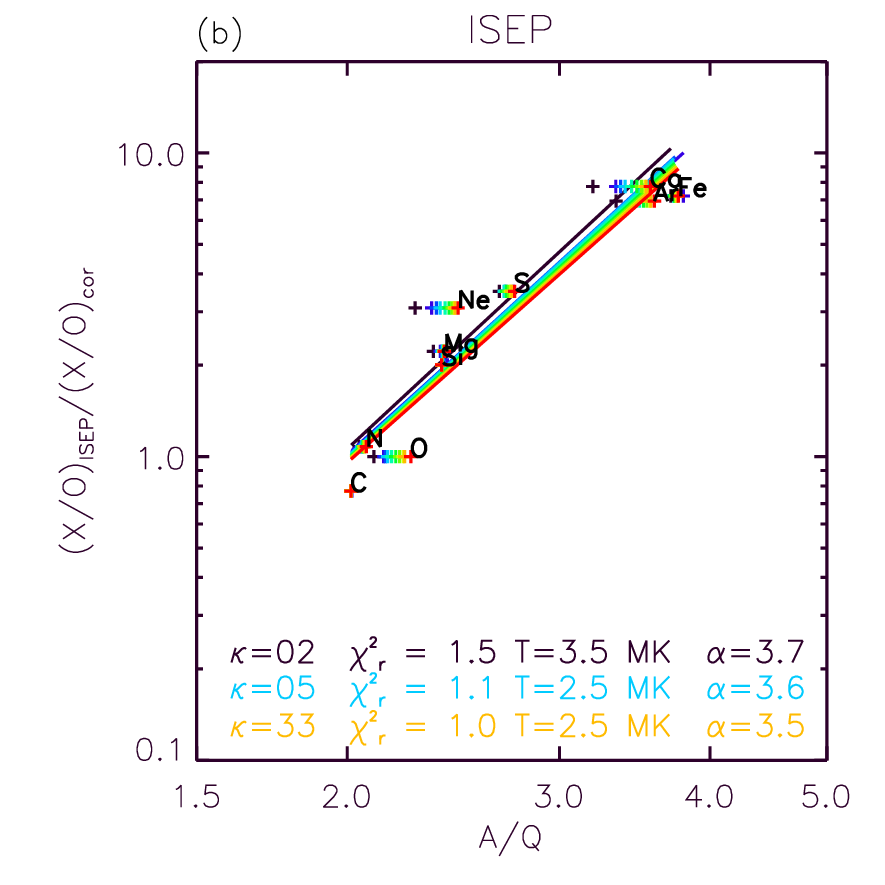}
\caption{(a)Reduced $\chi_r^{2}$ of the abundance enhancement power law fit  vs. temperatures (b)Abundance enhancement power law fit at the temperature of the second minimum of $\chi_r^{2}$. 
The fitting parameters are shown at $\kappa$=2, 5, and 33. See Table~\ref{tab:pwdi} for other $\kappa$ values.} 
\label{fig:chi}
\end{figure}

We find three $\chi_r^2$ minima in the temperature range of $10^5\sim10^8$~K (Figure~\ref{fig:chi}~(a)). Both the $\chi_r^2$ minima at coronal and transition region temperatures are statistically acceptable, but the higher temperature is not. 
The power-law fits at the second minimum are shown in Figure~\ref{fig:chi}~(b). This is a fit for a single temperature, and the situation where the SEPs are produced may be more complicated. 
A power law fit using a set of impulsive SEP abundances in \citet{reames14b} shows that the temperatures are 2.5$-$3.2~MK. 
These are similar to 2.51 MK at the second minimum of $\chi_r^2$ with various $\kappa$ values and 3.55~MK for $\kappa$ =2 (Table~\ref{tab:pwdi}).
The $\alpha$ is in the range of 3.52$-$3.62 with various $\kappa$ values and 3.52$\pm$0.36 with Maxwellian (Table~\ref{tab:pwdi}), 
and it is close to the the value, 3.64$\pm$0.15, in \citet{reames14b}.

At the first minimum at about 0.1$-$0.2~MK, the $\alpha$ is smaller at about 1.7$-$2.0. 
Cooler material, such as a filament near a jet eruption, could be a possible explanation for this lower temperature minimum as reported by \citet{mason16}.
However, their work is for a rare type of $^3$He-rich SEP events with enormously enhanced values of the S/O ratio, and we use the mean enhancement of 111 impulsive events. Therefore, this low temperature minimum would be worth investigating with individual events as a separate work in the future.  

\subsection{Distributions of power-law index and temperature} \label{subsec:plawd}

\citet{reames15} compiled distributions of T and the power index of $A/Q$, 
refining their work by adding 11 cooler He-poor (He/O/47$<$0.4) and 4 hotter (Ne/C/0.374$\leq$
Si/C/0.360) impulsive SEP events to the original 111 impulsive SEP events.  
Nearly all best fit plasma temperatures T
lay in the range 2.0-4.0 MK, but the mean of the distribution of power-law index of $A/Q$ was 4.47$\pm$0.07,
steeper than what might be expected from Figure~8 in their paper. Only 9 of their 126 power-law indices of $A/Q$ lay outside the range of 2.0-7.0. 

\begin{figure}
\centering
\includegraphics[width=100mm]{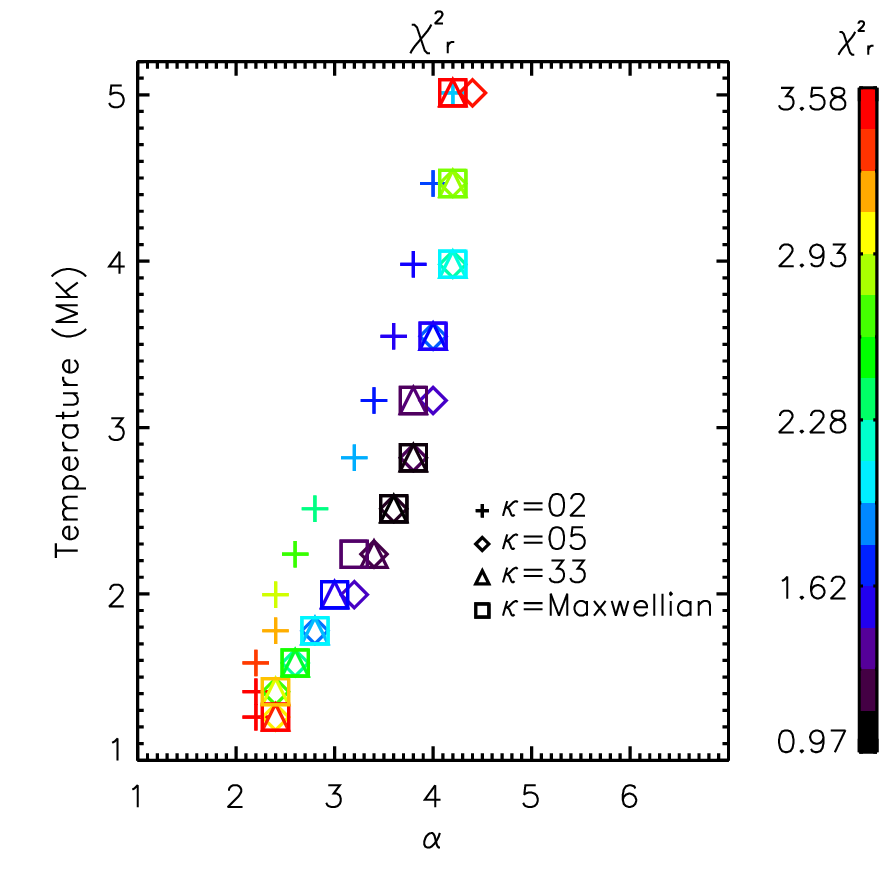}
\caption{The distributions of power-law index, $\alpha$ and temperature with $\kappa$=2, 5, and 33 and Maxwellian} 
\label{fig:pwd}
\end{figure}

We examine the distribution of the power-law index $\alpha$ and temperatures with the $\kappa$=2, 5, 33, and Maxwellian in Figure~\ref{fig:pwd}. 
The $\alpha$ is at around 4, and the temperature is 2$-$3~MK for the lower $\chi_{r}^{2}$ (close to black color). In the extreme case of $\kappa$=2, $\alpha$ is slightly higher, and the temperature is slightly higher (3$-$4 MK).   
Overall, we confirm that the temperatures and the power-law index values are within the range in previous works.  
Therefore, the derived source region temperature is not significantly affected by 
whether or not the electron velocity distribution deviates from a Maxwellian, i.e., thermal distribution. 

\section{Discussion} \label{sec:discussion} 

Impulsive SEP events are associated with small flares or coronal jets and type III radio bursts
\citep{reames20}, indicating magnetic reconnection allowing accelerated particles to escape along open magnetic fields to the
heliosphere \citep{bucik18a, bucik18b, bucik22}. The initial acceleration process is assumed to
occur in magnetic reconnecting current sheets, likely between a loop footpoint of one magnetic
polarity and adjacent open fields of opposite polarity \citep{reames21}. Extensive modeling of
electron acceleration during the reconnection process continues with increasingly sophisticated
models. Magnetic islands form in the current sheet and grow larger, reflecting electrons from
their ends in a classic Fermi acceleration process \citep{drake06}. Ion acceleration begins
with their entry into the reconnection exhausts governed by a threshold requirement favoring
high $A/Q$ ions for perpendicular heating \citep{drake12}. The main ion acceleration
takes place due to Fermi reflection in contracting and merging magnetic islands. The bulk of the
converted magnetic energy goes into forming a nonthermal electron power-law distribution, but
the electron energy gain can be suppressed by a strong guide field \citep{arnold21}, which may
initially be large, but then weaken as the flare progresses \citep{dahlin22}. 

Current modeling efforts assume constant ionic $A/Q$ values during the acceleration process \citep[e.g.][]{kramolis22}, 
so it is reasonable to suppose, as Reames does, that the best fit
temperatures derived from the SEP event abundance versus $A/Q$ plots reflect those of the SEP
seed populations. However, the electron acceleration proceeds with shorter time scales ($\simeq 10^{-3}$ s) than those of the ions ($\simeq$ 1 s) \citep{li22}, and that suggests that for sufficiently high plasma densities the nonthermal electron population may interact with the ions to enhance their charge states through ionization collisions. This could leave the ions with diminished $A/Q$ ratios through significant portions of the acceleration process. 

We have found that if the energy goes mainly into producing a $\kappa$ distribution with a high energy tail, the value of $A/Q$ is not strongly affected.  However, if the electrons are strongly heated, $A/Q$ can be decreased.  Both observations and models of the current sheets in solar flares show that Fe will be ionized from charge states near +13 at active region temperatures to charge states of at least +23 \citep{shen13, warren18}.  Since oxygen is already at a charge state of +6 to +7 in the active region, and it can only attain a charge state of +8 when fully stripped, its $A/Q$ can be only modestly reduced.  Therefore, one would expect a significant change in the Fe/O ratio.  It is possible that in the smaller jet events that produce impulsive SEPs, the SEPs leave the region of strong electron energization on a timescale shorter than the ionization time, which is around $10^9/n_e$ s.  It is also possible that electrons and ions are accelerated in different places.  However, our results suggest a third possibility -- that the electrons are accelerated into nonthermal distributions that do not drastically change $A/Q$. In particular, we note that the $\kappa$ distribution does not significantly improve $\chi^2$ over Maxwellian fits, so the observations do not require non-thermal distributions.

It is also important to note that elemental abundances may be the best way to determine the ionization state, and hence the temperature, in the region where SEPs are acclerated, because the ionization states measured in situ may have been altered by stripping as the SEPs propagate through the corona.

\section{Summary and Conclusion} \label{sec:conclusion}

A series of papers by Reames et al. has used the measured ionization states of SEPs and the assumption that the elemental abundances are enhanced in proportion to $(A/Q)^\alpha$ to determine the ionization equilibrium temperatures of the source regions.  The temperatures they found were typical active region temperatures, which is somewhat surprising in that electrons are also accelerated in the jets that produce impulsive SEPs, and solar flare plasma sheets show much higher charge states than the surrounding active regions.

We find that $\kappa$ distributions produce similar values of $A/Q$ to Maxwellian distributions with the same temperature, because at a given temperature the $\kappa$ distribution, as defined in Eq.(1),  shifts electrons to both higher and lower energies (Figure 1).  Therefore, one explanation for the observation that the source region temperatures derived for impulsive SEP events are typical active region temperatures is that electron energization produces a $\kappa$ distribution without greatly changing the temperature.  However, other explanations are also possible, including that electrons and ions are accelerated in different regions or that the SEPs leave the acceleration region in a time shorter than the ionization timescale.


\begin{acknowledgments}
Dr. Stephen Kahler had passed away during this work. We honor his dedication, passion for this work, and lifetime achievements in solar and space physics. 
CHIANTI is a collaborative project involving George Mason University, the University of Michigan (USA), University of Cambridge (UK) and NASA Goddard Space Flight Center (USA). This work was supported by the Air Force Office of Scientific Research under
award number FA2386-20-1-4031, the National Research Foundation of Korea(NRF) grant
funded by the Korea government(MSIT) (No. NRF-2020R1I1A1A01071814 and NRF-2023R1A2C1004167), 
and by NASA Grants NNH18ZD001N and HSR80NSSC21K2044 to the Smithsonian Astrophysical Observatory. Y.-K. Ko has been supported by Basic Research Funds at the Office of Naval Research.

\end{acknowledgments}


\bibliography{SEP-kappa}
\bibliographystyle{aasjournal} 

\begin{deluxetable}{ccccccc} 
\tabletypesize{\scriptsize}
\tablecaption{The mean abundance enhancement of the impulsive events} 
\tablehead{
\colhead{Element (X)} & \colhead{Z} & \colhead{ISEP$^a$} & \colhead{(X/O)$_{ISEP}$ } & \colhead{Coronal Abundance$^b$} & \colhead{(X/O)$_{cor}$} & \colhead{(X/O)$_{ISEP}$/(X/O)$_{cor}$} }
\startdata
He & 2  & 41400$\pm$2200 & 41.4 & 0.07943 & 102.329 & 0.404 \\
C & 6 & 386$\pm$8  & 0.386 & 0.00038 & 0.501 & 0.770 \\
N  & 7 &  139$\pm$4 & 0.139 & 0.00010 & 0.128 & 1.078  \\
O  & 8 &  1000$\pm$10 & 1.000 & 0.00077 & 1.0 & 1.000  \\
Ne & 10 &  478$\pm$24 & 0.478 & 0.00012 & 0.154 & 3.086  \\
Mg  & 12 &  404$\pm$30 & 0.404 & 0.00014 & 0.181 & 2.220  \\
Si  & 14 &  325$\pm$12 & 0.325 & 0.00012 & 0.162 & 2.003  \\
S  & 16 &  84$\pm$4 & 0.084 & 1.86$\times$10$^{-5}$ & 0.023 & 3.501  \\
Ar  & 18 &  34$\pm$2 & 0.034 & 3.80$\times$10$^{-6}$ & 0.004 & 6.941  \\
Ca  & 20 &  85$\pm$4 & 0.085 & 8.51$\times$10$^{-6}$ & 0.010 & 7.752  \\
Fe  & 26 &  1170$\pm$48 & 1.170 & 0.00012 & 0.162 & 7.214  \\
\enddata 
\label{tab:abundance}
\tablecomments{\\ 
$^a$ \citet{reames14b} \\
$^b$ Coronal elemental abundances relative to Hydrogen (Sun$\_$Coronal$\_$1992$\_$feldman$\_$ext.abund in CHIANTI, \citet{feldman92, grevesse98, landi02}) \\
 }
\end{deluxetable}

\begin{deluxetable}{ccccc} 
\label{tab:pwdi}
\tabletypesize{\scriptsize}
\tablecaption{The power law fitting parameters at the second minimum} 
\tablehead{
\colhead{ $\kappa$} & \colhead{$\chi^2_r$} & \colhead{Temperature (MK)} & \colhead{$\alpha$} & \colhead{Y-intercept} }
\startdata
2 & 1.49  & 3.55 & 3.58$\pm$0.39 & -1.08$\pm$0.16 \\
3 & 1.24  & 2.51& 3.56$\pm$0.37 & -1.07$\pm$0.16  \\
4 & 1.11  & 2.51 & 3.62$\pm$0.37 & -1.09$\pm$0.16  \\
5 & 1.06  & 2.51 & 3.62$\pm$0.37 & -1.09$\pm$0.16  \\
7 & 1.01  & 2.51 & 3.60$\pm$0.37 & -1.09$\pm$0.16 \\
10 & 0.98  & 2.51 & 3.58$\pm$0.37 & -1.09$\pm$0.16  \\
15 & 0.97  & 2.51 & 3.56$\pm$0.37 & -1.09$\pm$0.16  \\
25 & 0.97  & 2.51 & 3.54$\pm$0.36 & -1.08$\pm$0.16  \\
33 & 0.98  & 2.51& 3.53$\pm$0.36 & -1.08$\pm$0.16  \\
Maxwellian & 1.00 & 2.51 & 3.52$\pm$0.36 & \-1.08$\pm$0.16 
\enddata 
\end{deluxetable}



\end{document}